\RequirePackage[mathlines]{lineno} 
\documentclass[twocolumn,showpacs,aps,prl,amsmath,amssymb]{revtex4-1}
\usepackage[linktoc=all]{hyperref}
\usepackage{epsfig}
\usepackage{graphicx}
\usepackage{dcolumn}
\usepackage{bm}
\usepackage{overpic}
\usepackage{subfigure}
\usepackage{float}
\usepackage{color}
\usepackage{amsmath}
\usepackage{mathcomp}
\usepackage[center]{titlesec}  

\newcommand{\Dp}{\ensuremath{D^+}~}

\newcommand{\Do}{\ensuremath{D^0}~}

\newcommand{\Dpsig}{$D^+\to\pi^-\pi^+ e^+\nu_e$~}
\newcommand{\DPsig}{$D^+\to\pi^-\pi^+ e^+\nu_e$}
\newcommand{\Dosig}{$D^0\to\pi^-\pi^0 e^+\nu_e$~}

\begin{document}

\title{ \boldmath 
Observation of $D^+ \to f_0(500) e^+\nu_e$ and Improved Measurements of $D \to\rho e^+\nu_e$
}

\author{
M.~Ablikim$^{1}$, M.~N.~Achasov$^{10,d}$, S.~Ahmed$^{15}$, M.~Albrecht$^{4}$, M.~Alekseev$^{55A,55C}$, A.~Amoroso$^{55A,55C}$, F.~F.~An$^{1}$, Q.~An$^{52,42}$, Y.~Bai$^{41}$, O.~Bakina$^{27}$
% ...
\\
\vspace{0.2cm}
(BESIII Collaboration)\\
\vspace{0.2cm} {\it
$^{1}$ Institute of High Energy Physics, Beijing 100049, People's Republic of China\\
% ...
$^{4}$ Bochum Ruhr-University, D-44780 Bochum, Germany\\
% ...
$^{10}$ G.I. Budker Institute of Nuclear Physics SB RAS (BINP), Novosibirsk 630090, Russia\\
% ...
$^{15}$ Helmholtz Institute Mainz, Johann-Joachim-Becher-Weg 45, D-55099 Mainz, Germany\\
% ...
$^{27}$ Joint Institute for Nuclear Research, 141980 Dubna, Moscow region, Russia\\
% ...
$^{41}$ Southeast University, Nanjing 211100, People's Republic of China\\
$^{42}$ State Key Laboratory of Particle Detection and Electronics, Beijing 100049, Hefei 230026, People's Republic of China\\
% ...
$^{52}$ University of Science and Technology of China, Hefei 230026, People's Republic of China\\
% ...
$^{55}$ (A)University of Turin, I-10125, Turin, Italy; (B)University of Eastern Piedmont, I-15121, Alessandria, Italy; (C)INFN, I-10125, Turin, Italy\\
% ...
$^{d}$ Also at the Novosibirsk State University, Novosibirsk, 630090, Russia\\
% ...
}
}
\date{\today}

\begin{abstract}
Using a data sample corresponding to an integrated luminosity of 2.93~fb$^{-1}$ recorded by the BESIII detector at a center-of-mass energy of $3.773$ GeV,
we present an analysis of the decays \Dosig and \DPsig.
By performing a partial wave analysis, the $\pi^+\pi^-$ $S$-wave contribution to \Dpsig is observed to be $(25.7\pm1.6\pm1.1)\%$
with a statistical significance greater than 10$\sigma$, besides the dominant $P$-wave contribution.
This is the first observation of the $S$-wave contribution.
We measure the branching fractions
$\mathcal{B}(D^{0} \to \rho^- e^+ \nu_e) = (1.445\pm 0.058 \pm 0.039) \times10^{-3}$,
$\mathcal{B}(D^{+} \to \rho^0 e^+ \nu_e) = (1.860\pm 0.070 \pm 0.061) \times10^{-3}$, and
$\mathcal{B}(D^{+} \to f_0(500) e^+ \nu_e, f_0(500)\to\pi^+\pi^-) = (6.30\pm 0.43 \pm 0.32) \times10^{-4}$.
An upper limit of $\mathcal{B}(D^{+} \to f_0(980) e^+ \nu_e, f_0(980)\to\pi^+\pi^-) < 2.8 \times10^{-5}$ is set at the 90\% confidence level. 
We also obtain the hadronic form factor ratios of $D\to \rho e^+\nu_e$ at $q^{2}=0$ assuming the single-pole dominance parametrization:
$r_{V}=\frac{V(0)}{A_{1}(0)}=1.695\pm0.083\pm0.051$, $r_{2}=\frac{A_{2}(0)}{A_{1}(0)}=0.845\pm0.056\pm0.039$.
\\

\vspace{0.95cm}
\end{abstract}
\maketitle

The nature of the light scalar mesons $f_0(500)$, $f_0(980)$, and $a_0(980)$ has been controversial for many years~\cite{pdg}.
The investigation of their structure can improve our understanding of
the chiral-symmetry-breaking mechanisms of quantum chromodynamics (QCD) and quark confinement physics.
A $q\bar{q}$ configuration in the naive quark model cannot explain their mass ordering,
while there is still the possibility of being mixtures of $q\bar{q}$ states.
The other interpretations are often diquark-antidiquark states (tetraquark)~\cite{sigma-tetraquark}
and meson-meson bound states~\cite{sigma-molecular}.
The difficulty in unraveling this question has been due to the simultaneous presence of several different sources of nonperturbative strong interactions.

Since the leptons and hadrons in the final state interact with each other only weakly,
the semileptonic (SL) decay of $D^+ \to f_0(500) e^+\nu_e$ provides a unique and clean platform.
A sizable branching fraction (BF) of this decay is predicted by some theoretical models~\cite{Takayasu, YuJiShi}.
In addition, the $P$-wave dominance of the $\pi\pi$ system in this decay could be utilized to measure the hadronic form factor (FF),
which can in turn check theoretical approaches such as lattice QCD~\cite{LQCD} and QCD sum rules~\cite{QCDSR}.

In the previous study at the CLEO-c experiment~\cite{CLEO-D2rhoenu},
no significant indication for the $S$ wave was seen.
In this Letter, by performing a partial wave analysis (PWA) of \Dosig and \DPsig,
we report the first observation of $D^+ \to f_0(500) e^+\nu_e$, the measurements of the FF ratios for $D \to\rho e^+\nu_e$, and the related BFs.
For the BF measurement of SL decay, we use the double-tag technique~\cite{MarkIII-DTag}.
Charge conjugate states are implied throughout this Letter. 
The analysis is performed based on a data sample corresponding to an integrated luminosity of 2.93~$\rm fb^{-1}$~\cite{BESIII:2013iaa,BESIII:ISRPIPIPAPER} collected with the BESIII detector in $e^+e^-$ annihilation at a center-of-mass energy ($\sqrt{s}$) of $3.773$~GeV.
The BESIII detector is described in detail elsewhere~\cite{BESIII}.

The generic Monte Carlo (MC) sample, described in Ref.~\cite{bes3-Dp2kpienu},
has been verified to its validity to simulate the background in this analysis.
The signal MC sample consists of exclusive decays $\psi(3770)\to D\bar{D}$,
where the $D$ decays to the SL signal modes, with the decay-product distribution determined by the results of our PWA, while the $\bar{D}$ decays inclusively, as in the generic MC sample.

A detailed description of the selection criteria for charged and neutral particle candidates is provided in Ref.~\cite{bes3-Dp2kpienu}.
The tagged $\bar{D}$ mesons are reconstructed by appropriate combinations of the charged tracks and $\pi^0$ candidates in the following hadronic final states:
$K^{+}\pi^{-}$, $K^{+}\pi^{-}\pi^{0}$, $K^{+}\pi^{-}\pi^{0}\pi^{0}$, $K^{+}\pi^{-}\pi^{-}\pi^{+}$, and $K^{+}\pi^{-}\pi^{-}\pi^{+}\pi^{0}$ for neutral tags, 
and $K^{+}\pi^{-}\pi^{-}$, $K^{+}\pi^{-}\pi^{-} \pi^{0}$, $K_{S}^{0}\pi^{-}$, $K_{S}^{0}\pi^{-}\pi^{0}$, $K_{S}^{0}\pi^{-}\pi^{+}\pi^{-}$, and $K^{+}K^{-}\pi^{-}$ for charged tags.
The tag samples are selected based on two variables calculated in the $e^+e^-$ center-of-mass frame: $\Delta E \equiv E_{\bar{D}} - E_{\rm beam}$ and 
$M_{\rm BC} \equiv \sqrt{E_{\rm beam}^2 -  |\vec p_{\bar{D}}|^2}$,
where $E_{\bar{D}}$ and $\vec p_{\bar{D}}$ are the reconstructed energy and momentum of the $\bar{D}$ candidate, and $E_{\rm beam}$ is the beam energy.
If multiple candidates are present per tagged $\bar{D}$ mode, the one with the smallest $|\Delta E|$ is chosen. 
The yield of each tag mode is obtained from a fit to the $M_{\rm BC}$ distribution following Ref.~\cite{bes3-Dp2kpienu}.
We find (2759.6 $\pm$ 3.7)$\times 10^3$ and (1572.6 $\pm$ 1.5)$\times 10^3$ reconstructed neutral and charged tags, respectively.

After a tag is identified, we reconstruct the SL decay $D^{0(+)}\to\pi^-\pi^{0(+)} e^+\nu_e$ recoiling against the tag by
requiring an $e^+$ candidate and a $\pi^-\pi^{0(+)}$ pair following
Ref.~\cite{bes3-huangy}.
The momentum reconstruction of the $e^+$ candidate is improved by
recovering energy lost due to final-state radiation or bremsstrahlung
in the inner detector region.
If there are multiple $\pi^0$ candidates in an event, 
the $\gamma\gamma$ combination with its invariant mass closest to the nominal $\pi^0$ mass~\cite{pdg} is chosen.
To suppress the background to the $D^+$ signal from the decay of $D^+\to K^0_S e^+\nu_{e},\ K^0_S\to\pi^+\pi^-$,
we veto events with a $\pi^+\pi^-$ invariant mass within $\pm 70$ MeV/$c^2$ of the nominal $K^0_S$ mass~\cite{pdg},
which eliminates about 98.3\% of such background.
The reconstruction of the tag and SL decay candidates must include all charged tracks in the event and satisfy charge conservation.
In addition, the maximum energy of extra photon candidates ($E_{\gamma, \text{max}}$), which are not used in the tag and SL decay reconstruction,
is required to be less than 0.25 GeV to suppress the background events with extra $\pi^0$. 

Finally, we define the variable $U_{\rm miss} \equiv E_{\rm miss}-|\vec{P}_{\rm miss}|$ to identify the SL decay,
which peaks at zero for the signal since the neutrino is undetected.
Here $E_{\rm miss}$ and $\vec{P}_{\rm miss}$ are the missing energy and momentum of the $D$ meson;
they are calculated in the $e^+e^-$ center-of-mass frame by $E_{\rm miss}=E_{\rm beam}-E_{\rm \pi\pi}-E_{e}$ and $\vec{P}_{\rm miss}=\vec{P}_{\rm SL}-\vec{P}_{\pi\pi}-\vec{P}_{e}$,
where $E_{\pi\pi}$ and $\vec{P}_{\pi\pi}$ are the energy and momentum of $\pi\pi$ system, $\vec{P}_{\rm SL}$ is the momentum of the SL candidate,
which is calculated as $\vec{P}_{\rm SL}=-\hat{P}_{\rm tag}\sqrt{E_{\rm beam}^{2}-m_{\bar{D}}^{2}}$ to improve the $U_{\rm miss}$ resolution.
Here $\hat{P}_{\rm tag}$ denotes the unit momentum vector of the $\bar{D}$ tag and $m_{\bar{D}}$ is the nominal $\bar{D}$ mass~\cite{pdg}.

\begin{figure}[htp]
\begin{center}
\includegraphics[width=0.36\textwidth]{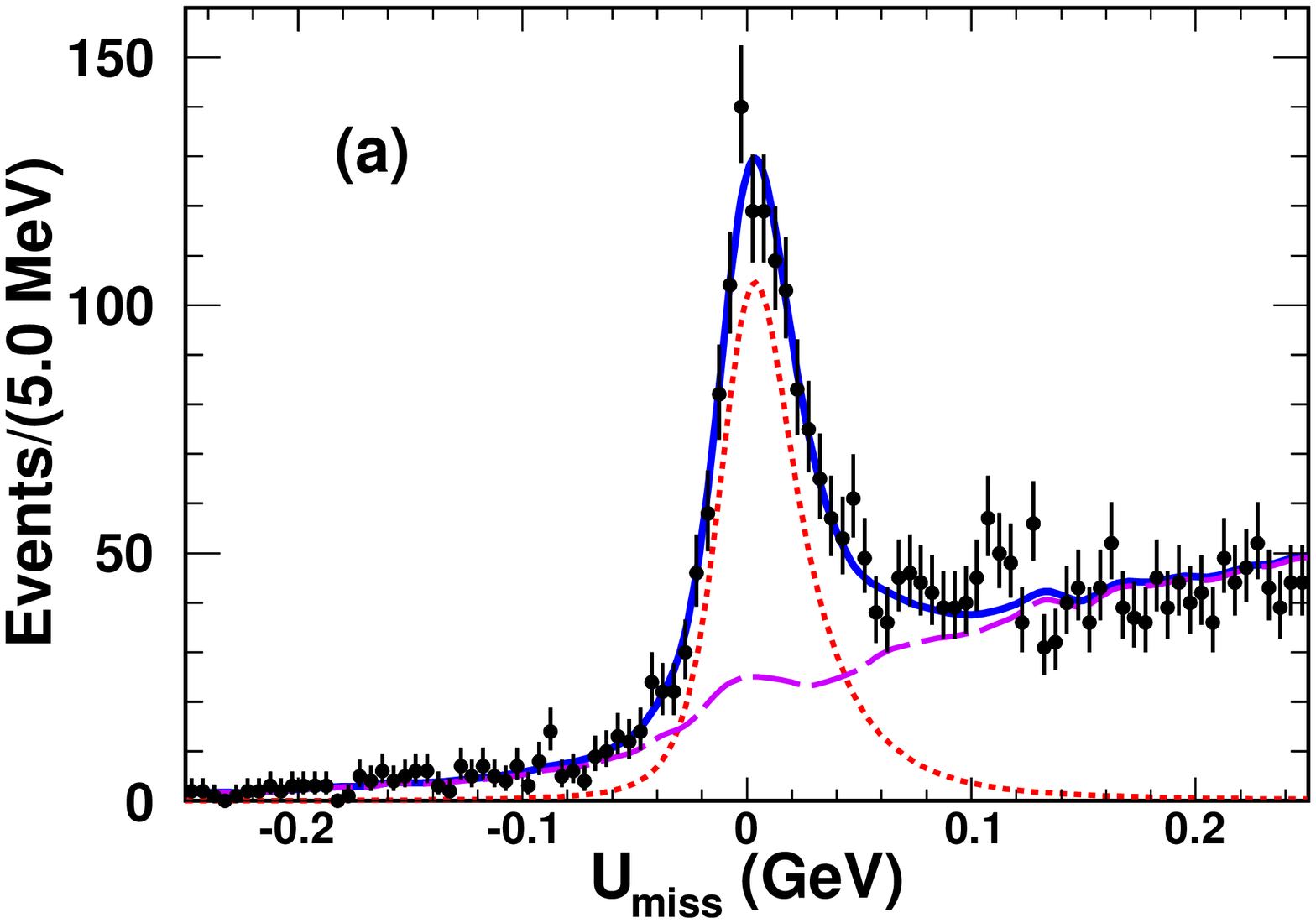}
\includegraphics[width=0.36\textwidth]{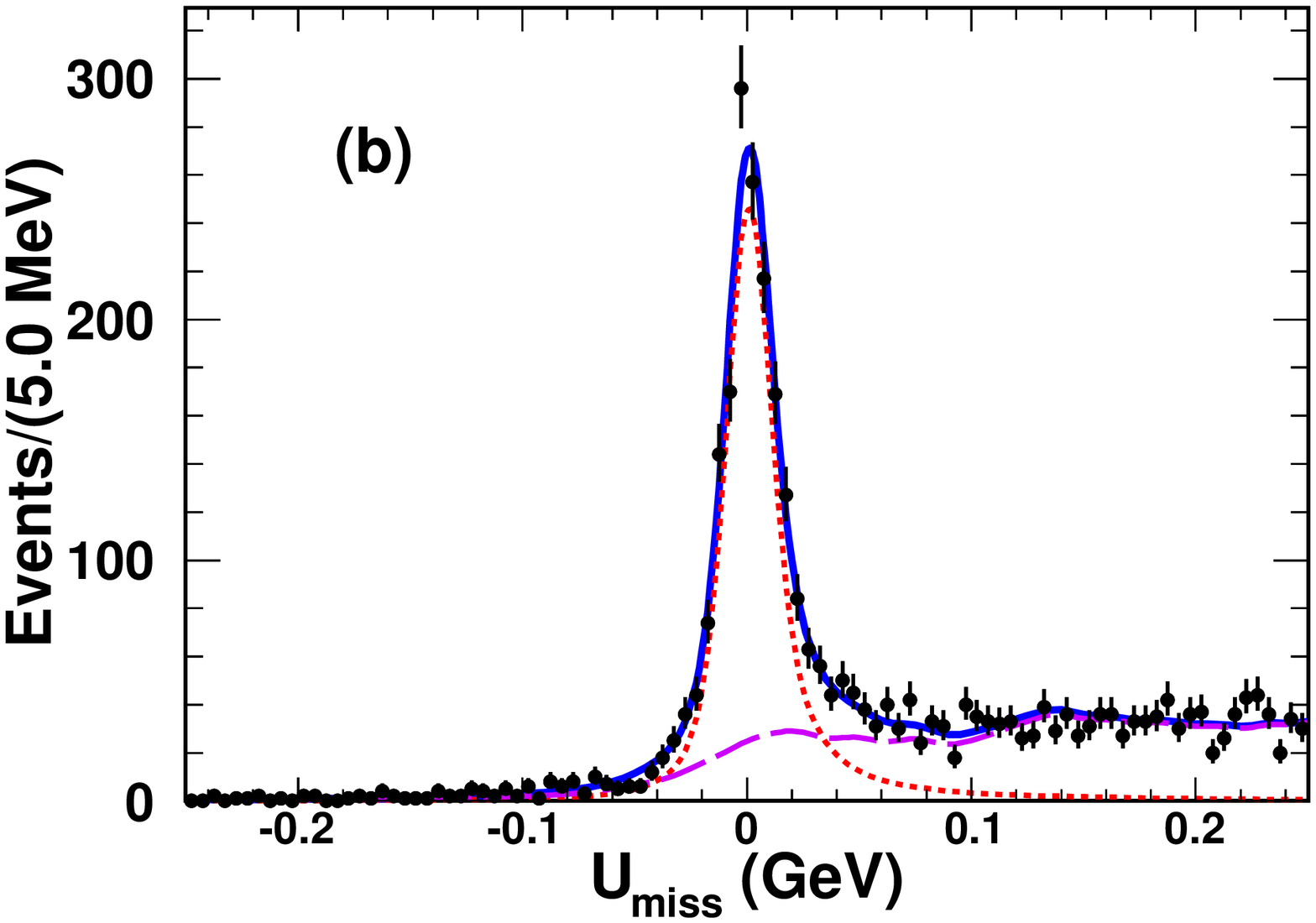}
\caption{ Fits to the $U_{\rm miss}$ distributions for \Dosig (a) and \Dpsig (b).
The points with error bars are data, and the solid lines are the fits. The short-dashed lines are signals and the long-dashed lines are backgrounds.
}
\label{fit:Umiss-data}
\end{center}
\end{figure}

The main background contributions are from $D\bar{D}$ decays, while backgrounds from other processes are negligible.
For the $D^0$ decay, the dominant background arises from $D^0\to K^*(892)^- e^+\nu_e$,
which results in $U_{\rm miss}$ distribution that is predominantly greater than zero.
The backgrounds that peak in $U_{\rm miss}$ mostly arise from $D^0\to K^{-}e^+\nu_e,~K^-\to\pi^-\pi^0$ and $D^+\to K^{0}_S e^+\nu_e$ decays.
For the $D^+$ decay, the background is dominated by
$D^+\to\bar{K}^*(892)^0 e^+\nu_e$, which peaks near zero and $\pi^0$
mass, depending on the $\bar{K}^*(892)^0$ decay mode.
With all tag modes combined, we extract the signal yields by performing an unbinned-maximum-likelihood fit to the $U_{\rm miss}$ distribution.
The signal is described by the signal MC distribution convolved with a Gaussian function,
and the background is modeled by the generic MC distribution convolved with the same Gaussian resolution function.
The mean and standard deviation of the Gaussian function are left free to account for any difference between the $U_{\rm miss}$ resolution in the MC simulation and the data.
The fit results are shown in Fig.~\ref{fit:Umiss-data}. We obtain signal yields of $1102\pm45$ and $1667\pm50$
for \Dosig and \DPsig, respectively, where the errors are statistical.

\begin{figure}[htp]
\includegraphics[width=0.195\textwidth]{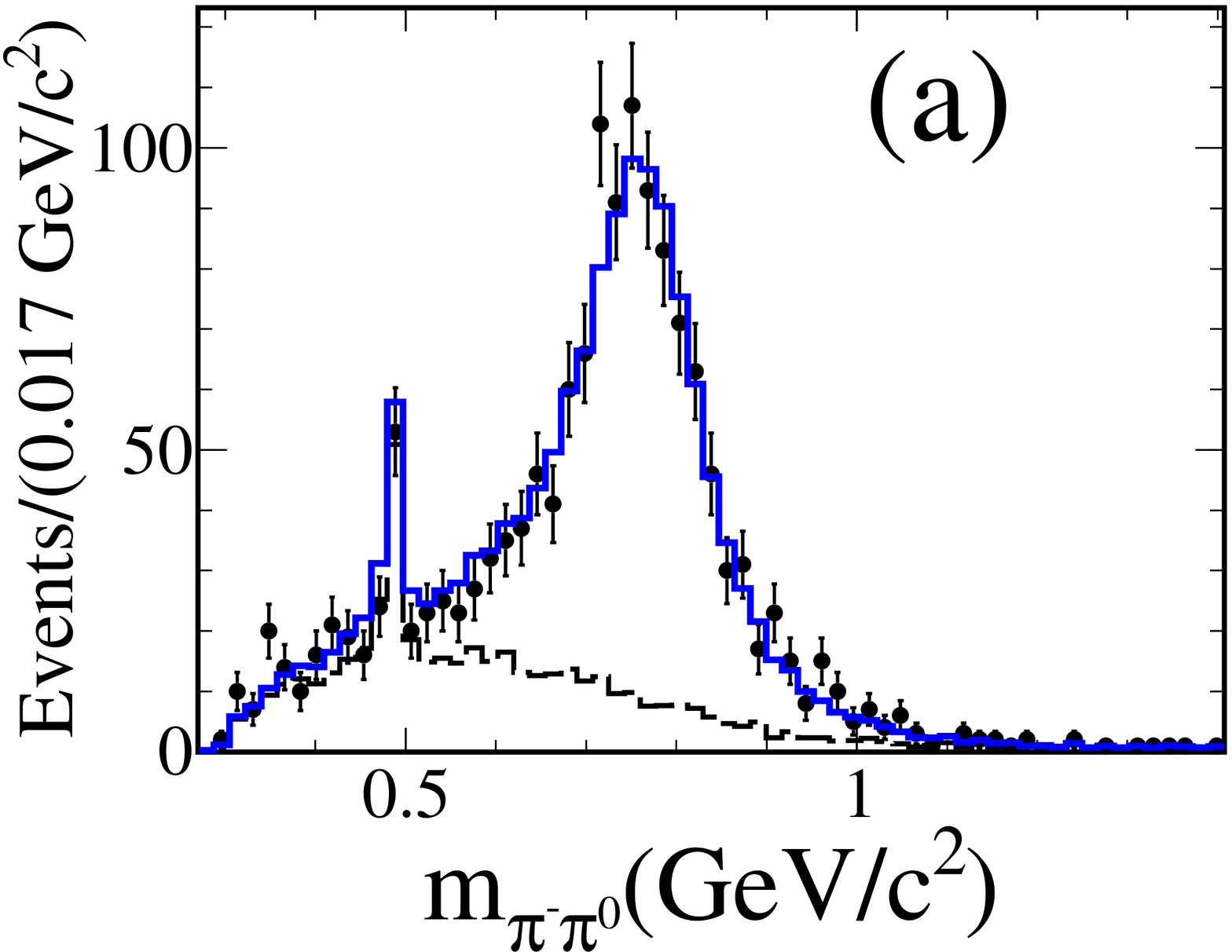}
\includegraphics[width=0.195\textwidth]{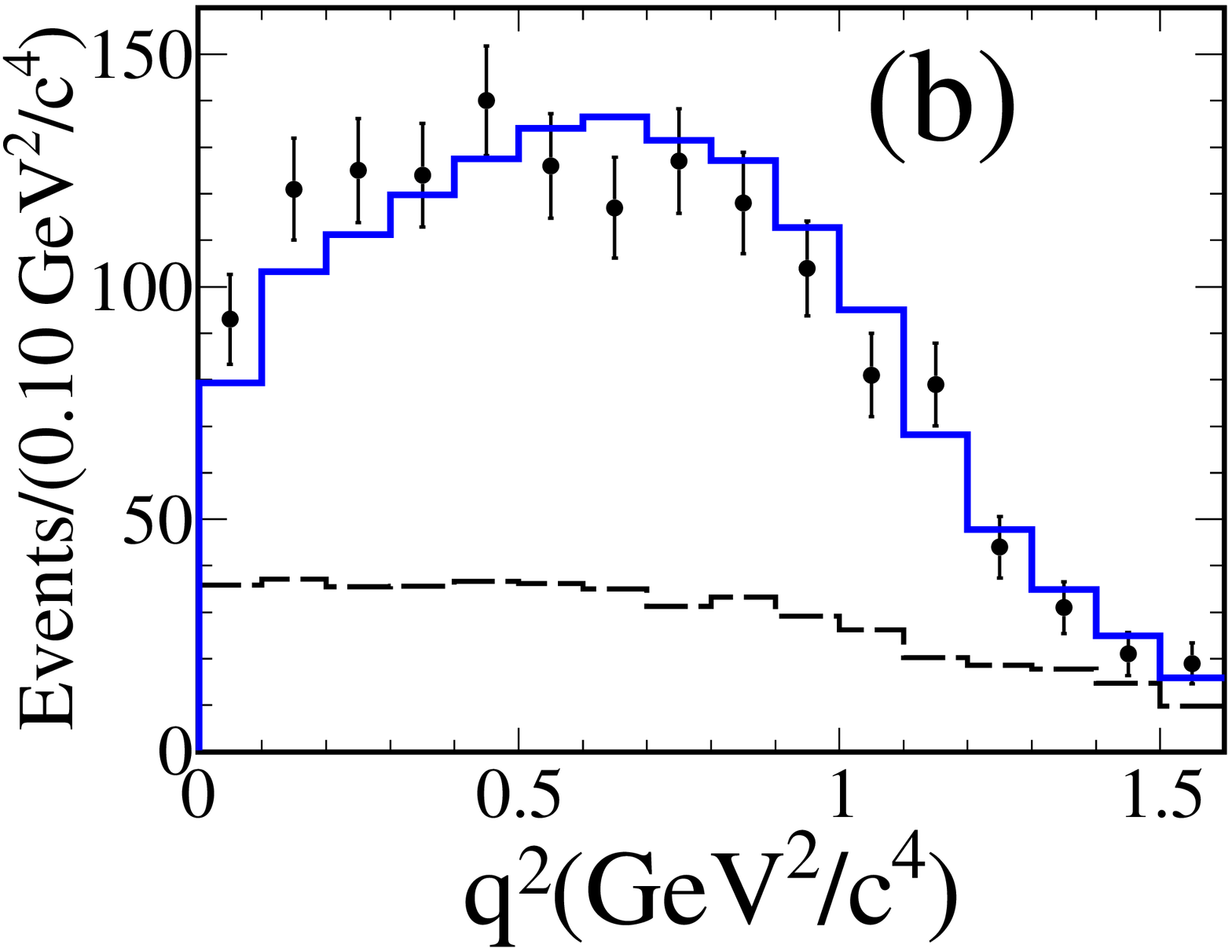} 
\includegraphics[width=0.195\textwidth]{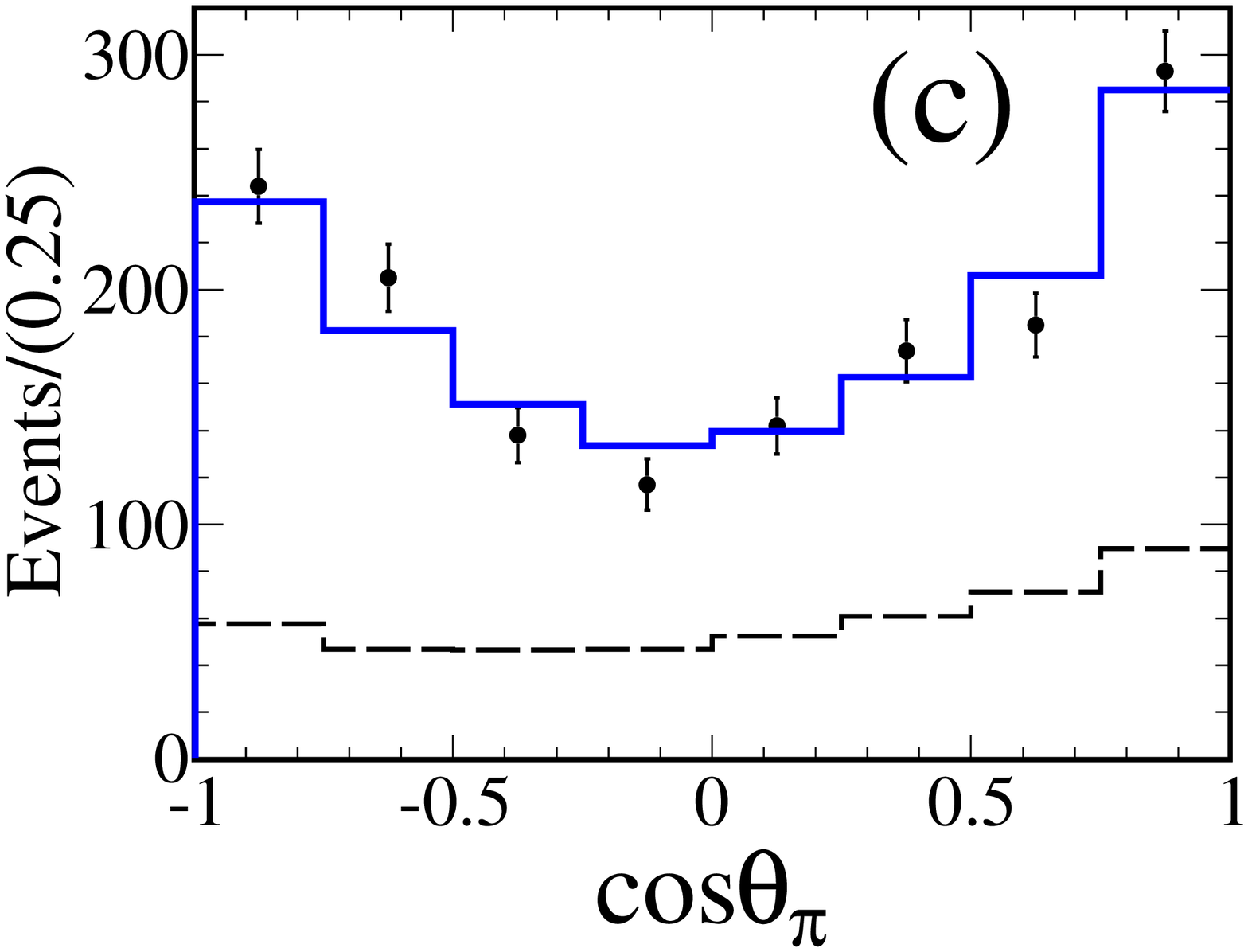}
\includegraphics[width=0.195\textwidth]{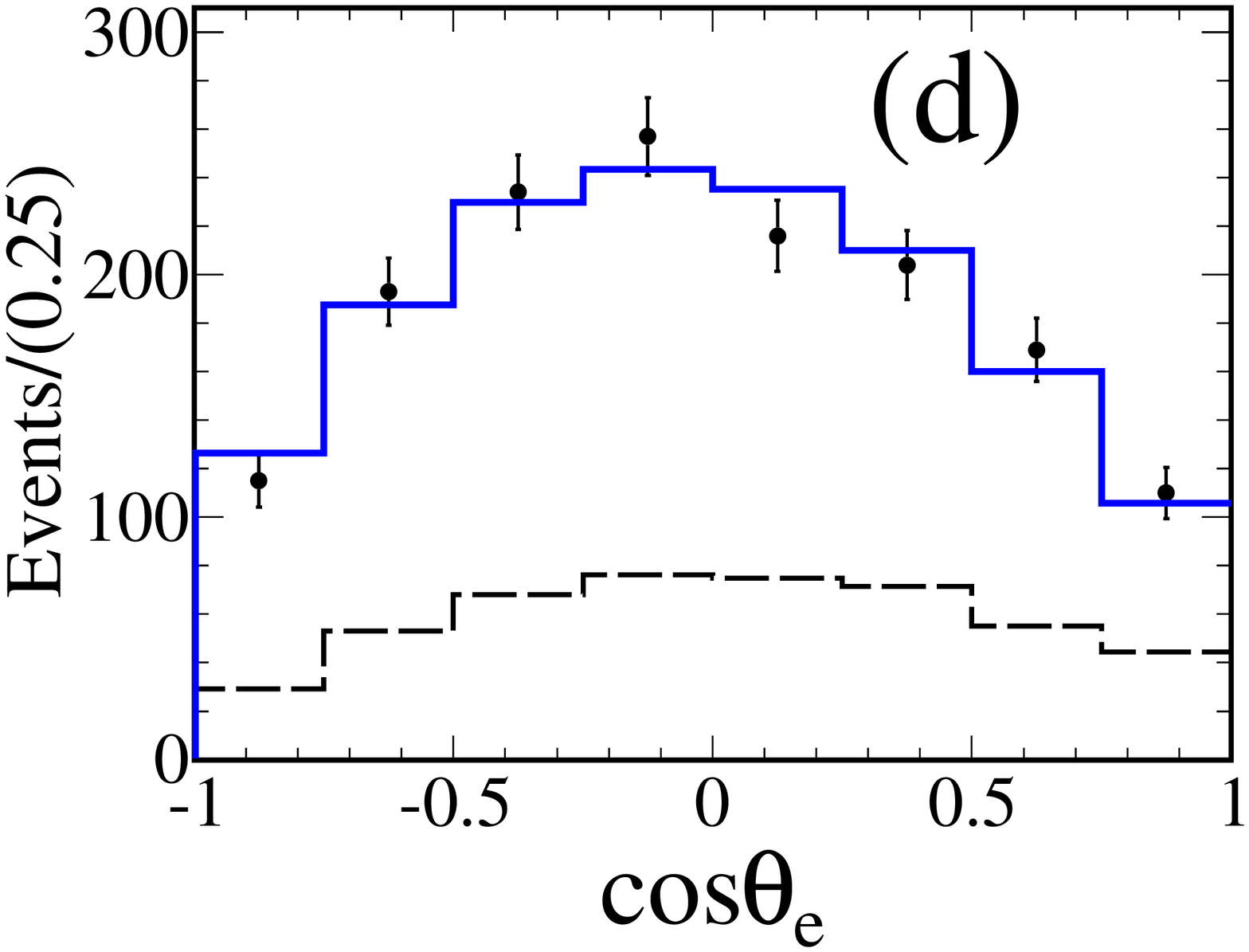}
\includegraphics[width=0.195\textwidth]{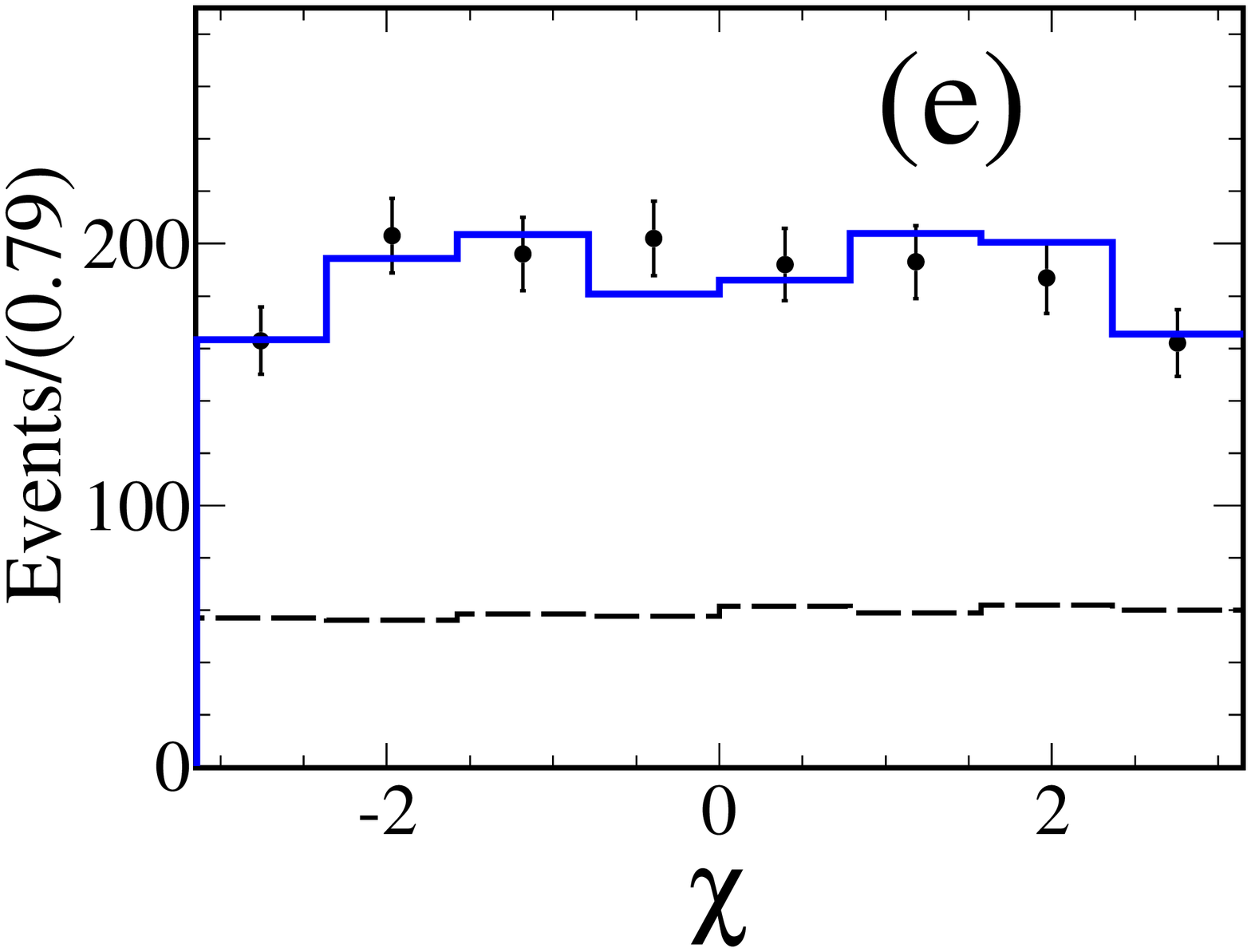}\\
\includegraphics[width=0.195\textwidth]{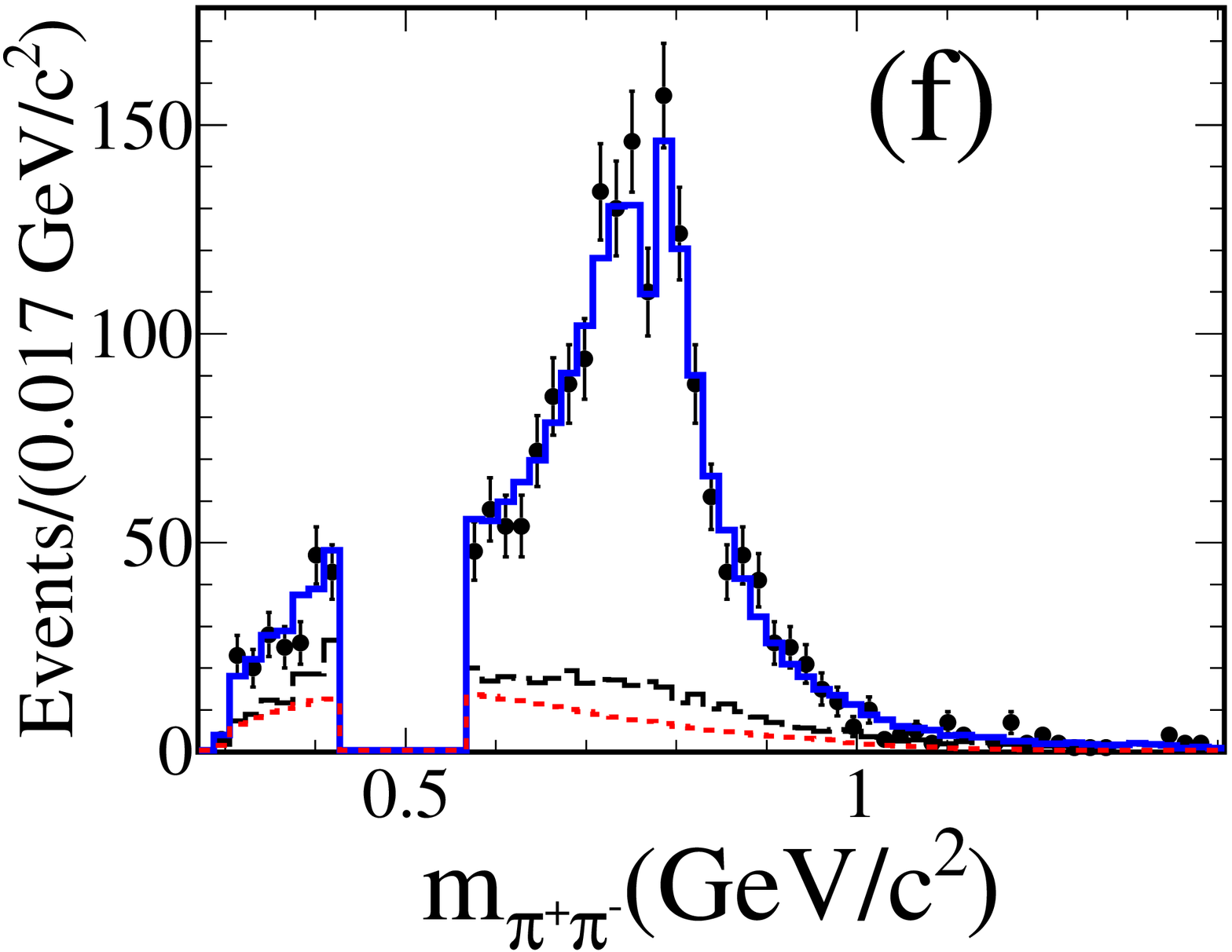}
\includegraphics[width=0.195\textwidth]{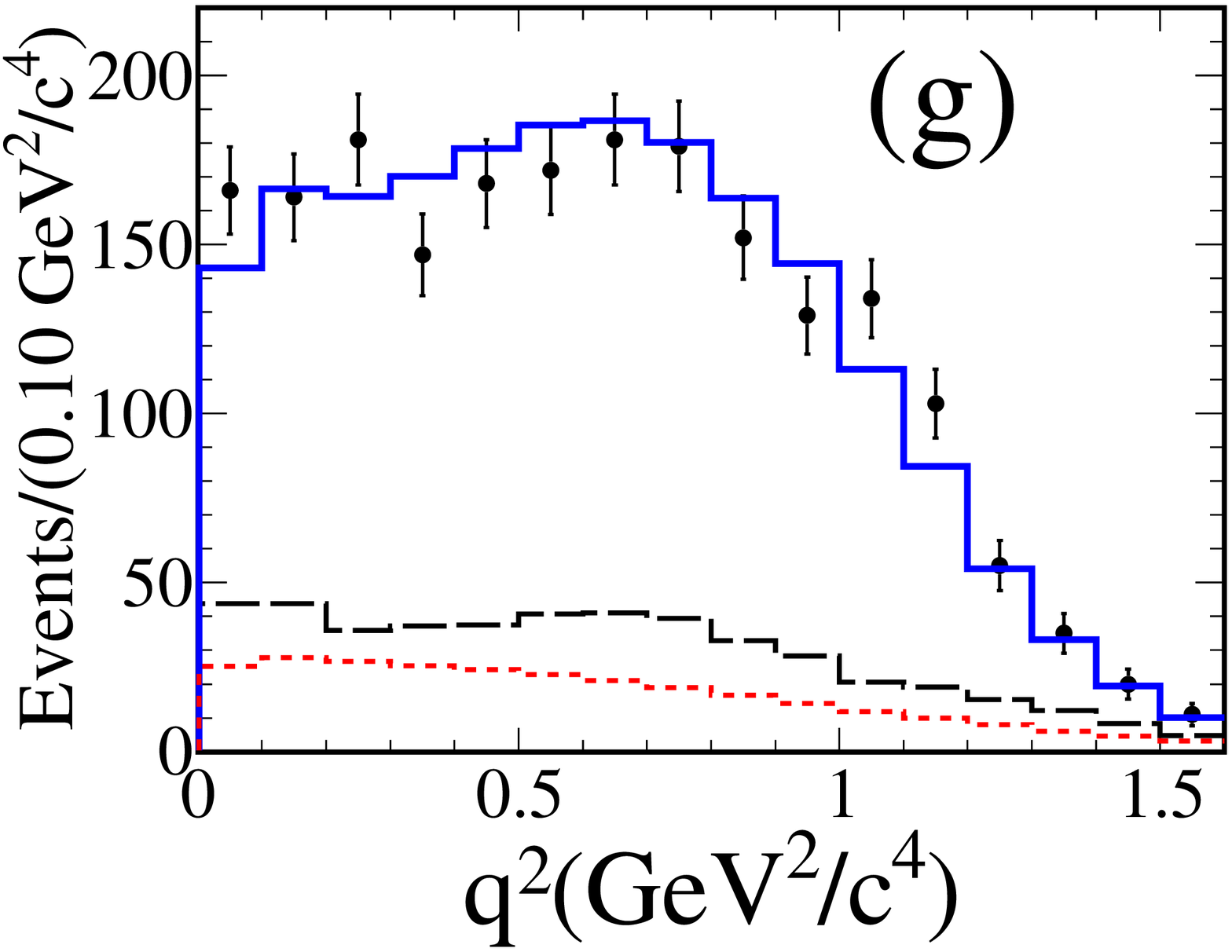} 
\includegraphics[width=0.195\textwidth]{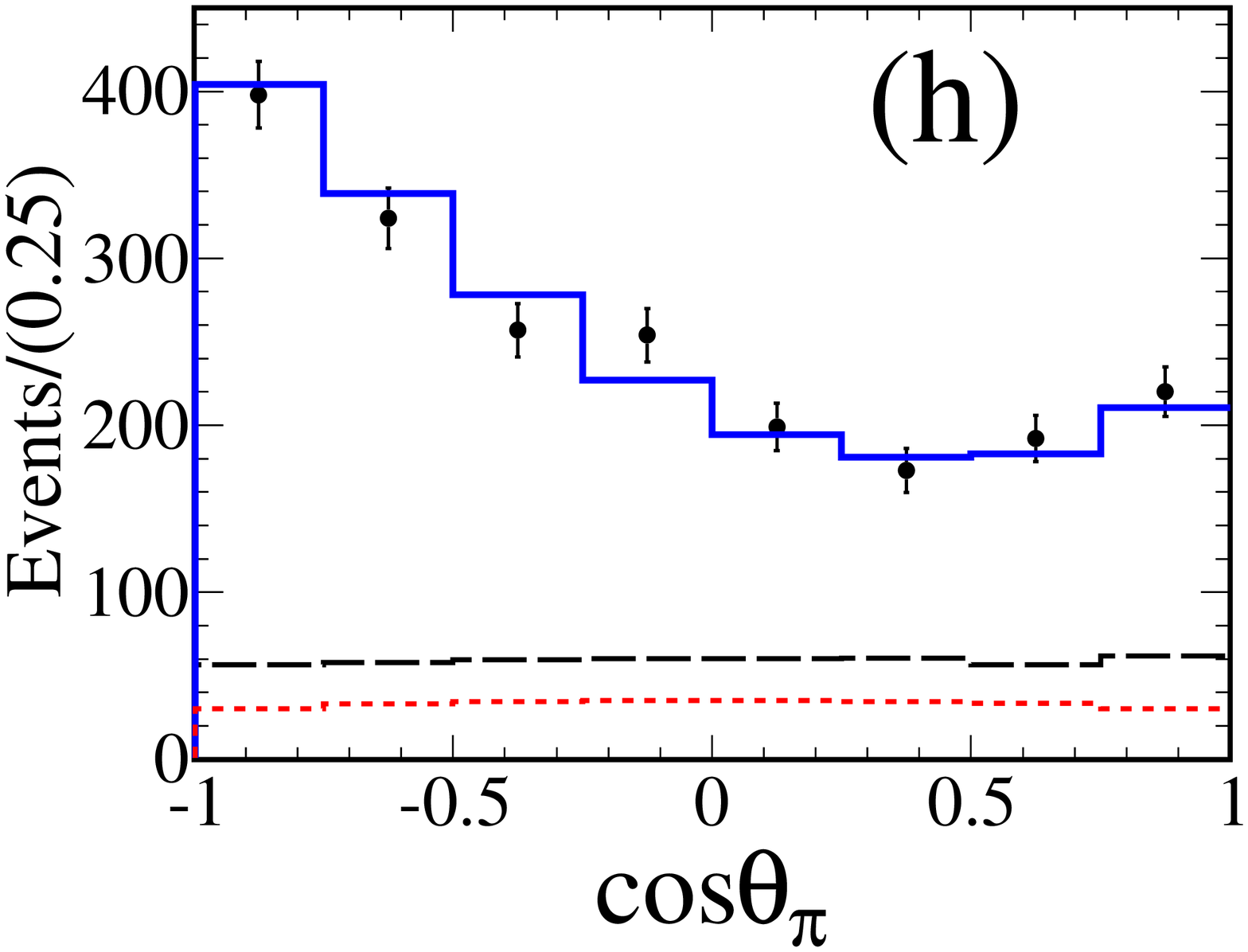}
\includegraphics[width=0.195\textwidth]{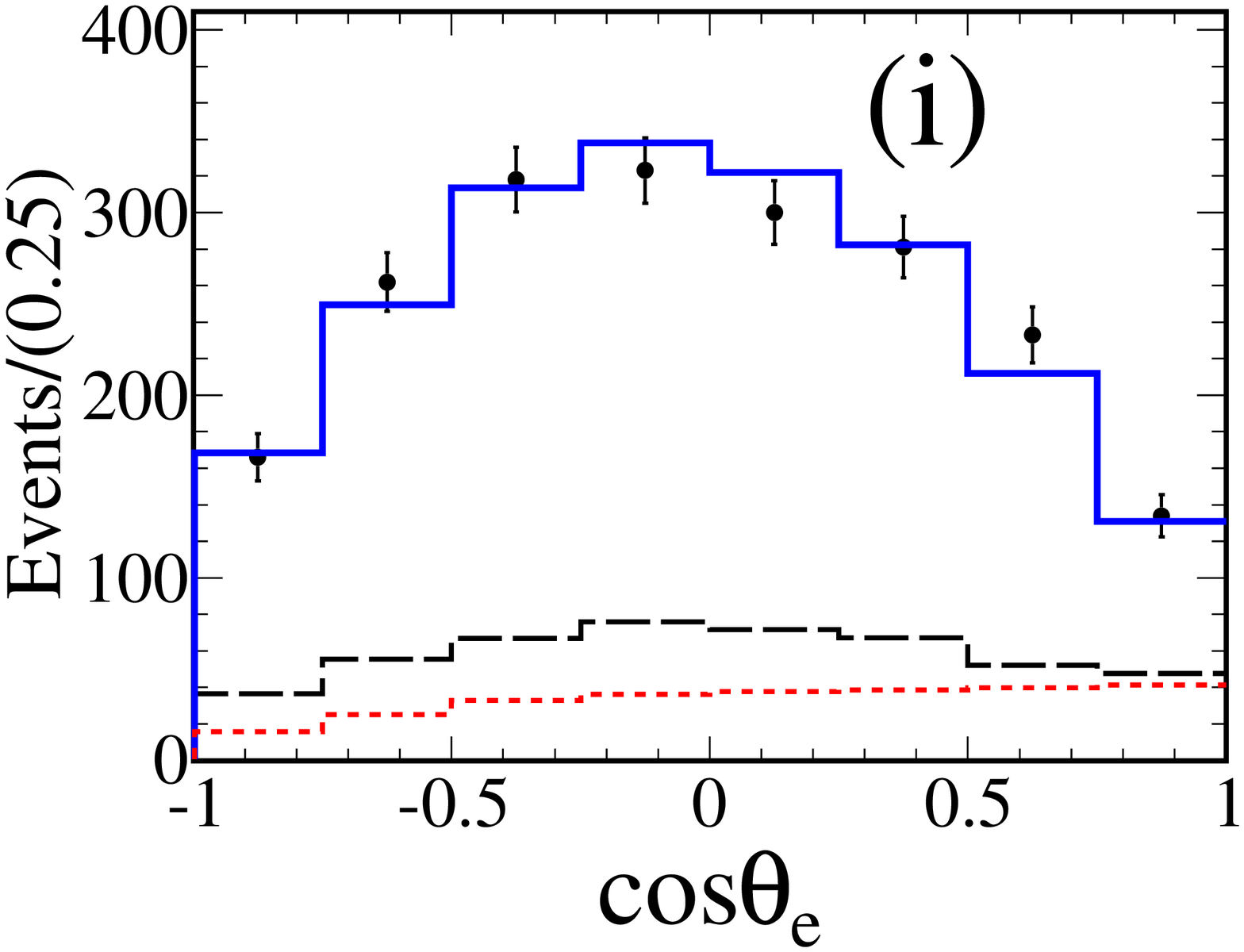}
\includegraphics[width=0.195\textwidth]{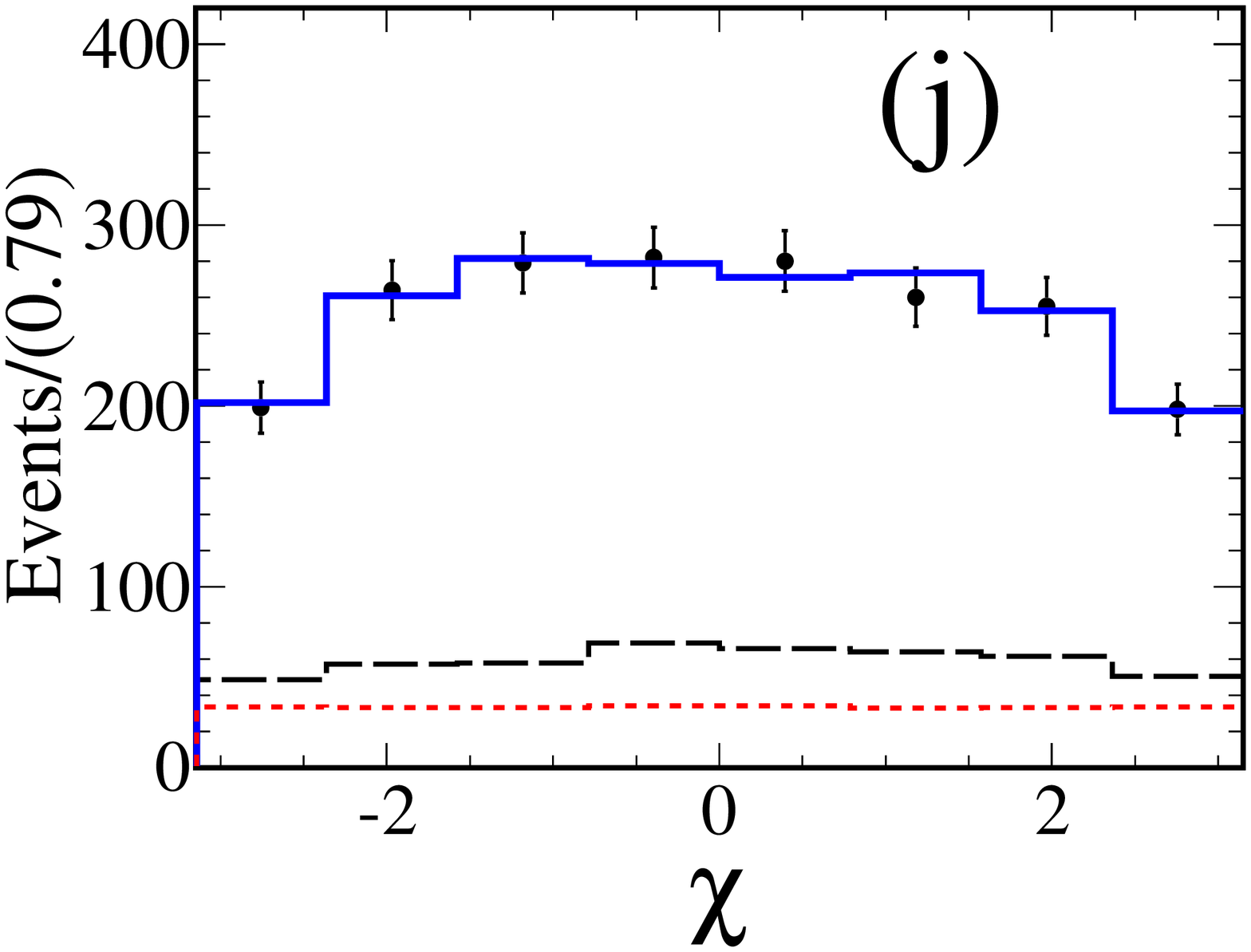}
\caption{Projections of the data and simultaneous PWA fit onto the five kinematic variables for \Dosig (top) and \Dpsig (bottom) channels.
The dots with error bars are data, the solid lines are the fits, the dashed lines show the MC simulated backgrounds,
and the short-dashed lines in (f)-(j) show the component of $D^+\to f_0(500)e^+\nu_e$.
}
\label{fit:double-D}
\end{figure}

To study the $\pi\pi$ system and measure the FF, we require $|U_{\rm miss}|< 0.06$ GeV to select samples for PWA;
this leads to 1498 [2017] events with a background fraction of $(33.28\pm0.87)$\% [$(23.82\pm0.69)$\%] in the $D^0$ [$D^+$] mode.
The differential decay rate for $D^{0(+)} \rightarrow \pi^- \pi^{0(+)} e^+ \nu_e$ depends on five variables \cite{ref:cab1, BaBar-Dp2kpienu}:
$m$, the invariant mass of the $\pi\pi$ system;
$q$, the invariant mass of the $e^+\nu_e$ system;
$\theta_e$ ($\theta_{\pi}$), the angle between the momentum of the $e^+$ ($\pi^-$) in the $e^+\nu_e$($\pi\pi$) rest frame and
the momentum of the $e^+\nu_e$ ($\pi\pi$) system in the $D$ rest frame;
and $\chi$, the angle between the normals of the decay planes defined in the $D$ rest frame by the $\pi\pi$ pair and the $e^+\nu_e$ pair.
The sign of $\chi$ should be changed when analyzing a $\bar{D}$ candidate in order to maintain $CP$ conservation.
In theory, the differential decay rate as a function of these variables is given in Ref.~\cite{ref:wise1}.
Neglecting the contributions from the positron mass, it depends on the hadronic FFs as defined in Ref.~\cite{BaBar-Dp2kpienu}.
For the $P$-wave contribution, we use the Gounaris-Sakurai (GS) function~\cite{GS} to describe $\rho^-$ and $\rho^0$;
the $\rho^0-\omega$ interference is taken into account by the form
$R_{\rho^0-\omega}(m) = \textrm{GS}_{\rho^0}(m) \times [1 + a_{\omega}e^{i\phi_{\omega}}  \textrm{RBW}_{\omega}(m)]$,
where RBW is a relativistic Breit-Wigner function with a constant width~\cite{rhoomega}.
A Blatt-Weisskopf damping factor $(r_{\rm BW})$ related to the meson radii is included in the decay amplitude.
The $q^2$ dependence of the total FFs are parametrized in terms of one vector FF $[V(q^2)]$ and two axial vector FFs $[A_{1,2}(q^2)]$
that are assumed to be dominated by a single pole: 
$V(q^2)=\frac{V(0)}{1-q^2/m_V^2},~A_{1,2}(q^2)=\frac{A_{1,2}(0)}{1-q^2/m_{A}^2}$.
Here $m_{V}$ and $m_{A}$ are the pole masses and fixed to $m_{D^{*}(1^-)}\simeq$ 2.01 GeV/$c^{2}$
and $m_{D^{*}(1^+)}\simeq$ 2.42 GeV/$c^{2}$~\cite{pdg} in the fit, respectively.
At $q^2=0$, the FF ratios, $r_{V}=\frac{V(0)}{A_{1}(0)}$ and $r_{2}=\frac{A_{2}(0)}{A_{1}(0)}$, are determined from the fit to the differential decay rate.
These ans\"{a}tze are adequate according to the fit results shown in Figs.~\ref{fit:double-D} (b) and ~\ref{fit:double-D} (g).
The $S$-wave contribution, characterized by the FF ${\cal F}_{10}$, is parametrized, assuming only $f_0(500)$ production, as 

\begin{equation}
{\cal F}_{10}=p_{\pi\pi}m_{D}\frac{a_{S}e^{i\phi_S}\mathcal{A}_{S}(m)}{1-\frac{q^{2}}{m_{A}^{2}}},
\label{eq:form_factor_S}
\end{equation}
where $p_{\pi\pi}$ is the magnitude of the three-momentum of the $\pi\pi$ system in the $D$ rest frame.
Here the term $\mathcal{A}_{S}(m)$ corresponds to the mass-dependent $S$-wave amplitude modeled by the fixed resonant line shape described in Ref.~\cite{Bugg};
the parameters $a_{S}$ and $\phi_S$ are the magnitude and phase of $\mathcal{A}_{S}(m)$ relative to $\textrm{GS}_{\rho^0}(m)$.

We perform the PWA using an unbinned-maximum-likelihood fit.
The negative log likelihood $- \ln {\mathcal L}$ is defined as

\begin{equation} 
-\sum_{i=1}^{N}\ln \left((1-f_b)\frac{\omega(\xi_{i},\eta)}{\int d\xi_{i}\omega(\xi_{i},\eta)\epsilon(\xi_i)} + f_b\frac{B_{\epsilon}(\xi_{i})}{\int d\xi_i B_{\epsilon}(\xi_i)\epsilon(\xi_i)}\right),
\label{eq:nll}
\end{equation}
where $\xi_i$ denotes the five kinematic variables characterizing the $i{\rm th}$ event of $N$ and $\eta$ denotes the fit parameters;
$\omega(\xi_i,\eta)$ is the decay intensity, 
and $B_{\epsilon}(\xi_i)$ is defined to be the background distribution corrected by the acceptance function $\epsilon(\xi_i)$~\cite{CLEO-D02KKPiPi}.
The background shape is parametrized using the generic MC and its fraction $f_b$ is fixed according to the result of the $U_{\rm miss}$ fit.
We model the background with a nonparametric function class RooNDKeysPdf~\cite{bkgfun2}
that uses an adaptive kernel-estimation algorithm~\cite{bkgfun3}.
The normalization integral in the denominator is determined using a MC technique~\cite{bes3-Dp2kpienu}. 

A simultaneous PWA fit is performed on both isospin-conjugate modes. 
The structure of the $\pi\pi$ system is only the $\rho^-$ in the \Do mode and is dominated by the $\rho^0$, with a small fraction of $\omega$, in the \Dp mode. 
In the fit, the masses and widths of $\rho$ and $\omega$ are fixed to those reported in Ref.~\cite{pdg}.
We also consider other possible components in the \Dp mode, especially a $\pi^+\pi^-$ $S$-wave contribution from the $f_0(500)$. 
We find that the cos$\theta_{\pi}$ distribution of the fit can agree with data only after considering the $S$-wave contribution.
The statistical significance of the $f_0(500)$ is determined to be more than $10\sigma$
from the change of $-2\ln {\mathcal L}$ in the PWA fits with and without this component, taking into account the change of the number of degrees of freedom.
The projections of the five kinematic variables for the data are shown in Fig.~\ref{fit:double-D}.
The difference of the cos$\theta_{\pi}$ distribution between two modes is due to the $\pi^+\pi^-$ $S$-wave interference contribution in $D^+$ decays.
Based on this nominal solution, we obtain the fractions of the different components: $f_{f_0(500)}=(25.7\pm1.6\pm 1.1)\%$, $f_{\rho^{0}}=(76.0\pm1.7\pm 1.1)\%$
and $f_{\omega}  =(1.28\pm0.41\pm 0.15)\%$, as well as the FF ratios $r_V=1.695\pm0.083\pm0.051$ and $r_2=0.845\pm0.056\pm0.039$,
with a correlation coefficient $\rho_{r_V,r_2}=-0.206$,
where the first and second uncertainties are statistical and systematic, respectively.
To calculate the fractions and estimate the corresponding statistical uncertainties, we employ the same method described in Ref.~\cite{bes3-luy}.
As a cross check, we perform fits to the two modes separately, and the results are consistent with the simultaneous fit.

\onecolumngrid

\twocolumngrid

Replacing the $f_0(500)$ component with a phase-space $S$-wave amplitude worsens the $-\ln{\mathcal L}$ by 40.3.
If the phase-space $S$-wave amplitude is added to the nominal solution on top of the $f_0(500)$ component,
its statistical significance is only about 1$\sigma$, so this contribution is neglected.
In addition, a possible $f_0(980)$ component contributing to the ${\cal F}_{10}$ term is studied by adding it to the nominal solution,
where $f_0(980)$ is parametrized by the Flatt\'e formula with its parameters fixed to the BESII measurements~\cite{PLB607}. 
The significance of this component is less than 2$\sigma$.
By scanning the BF of the $f_0(980)$ component in the physical region,
we obtain an upper limit at the 90\% confidence level (CL), which is listed in Table~\ref{tab:Bf}.
To take the systematic uncertainty into account, the likelihood is convolved with a Gaussian function with a resolution equal to the systematic uncertainty.

We calculate the absolute BFs of both modes with the same method as described in Ref.~\cite{bes3-Dp2kpienu}.
For the $D^0$ mode, the only significant contribution observed is $D^0\to\rho^-e^+\nu_e$.
For the $D^+$ mode, the absolute BFs of the different components are derived from $\mathcal{B}(D^{+}\to\pi^{-}\pi^{+}e^{+}\nu_e)\times f_{i}$,
where $i$ denotes the different components of the $\pi\pi$ system: $f_0(500)$, $\rho^0$, and $\omega$, and $f_i$ denotes the fraction obtained via the PWA.
The BFs of $\pi^0\to\gamma\gamma$ and $\omega\to\pi^+\pi^-$~\cite{pdg} have been included in the calculation.
All the results are summarized in Table~\ref{tab:Bf}.

\begin{table}[htp] 
\begin{center}
\caption{Measured absolute BFs and upper limit of the BF at 90\% CL.
The first (second) uncertainties are statistical (systematic).
}
\vspace{0.15cm}
\scalebox{0.88}{
\begin{tabular}{lc}
\hline
\hline
Signal mode  &  this analysis ($\times 10^{-3}$)           \\
\hline
\vspace{-0.30cm}\\

$D^0\to\pi^-\pi^0 e^+\nu_{e}$         &  1.445 $\pm$ 0.058 $\pm$ 0.039   \\

$D^0\to\rho^- e^+\nu_{e}$             &  1.445 $\pm$ 0.058 $\pm$ 0.039   \\

$D^+\to\pi^-\pi^+e^+\nu_{e}$          &  2.449 $\pm$ 0.074 $\pm$ 0.073   \\

$D^+\to\rho^0 e^+\nu_{e}$             &  1.860 $\pm$ 0.070 $\pm$ 0.061   \\

$D^+\to\omega e^+\nu_{e}$             &  2.05 $\pm$ 0.66 $\pm$ 0.30      \\

$D^+\to f_0(500)e^+\nu_{e}, f_0(500)\to\pi^+\pi^-$  &  0.630 $\pm$ 0.043 $\pm$ 0.032  \\

$D^+\to f_0(980)e^+\nu_{e}, f_0(980)\to\pi^+\pi^-$  &  $ < 0.028$ \\
\hline
\hline
\end{tabular}
}
\label{tab:Bf}
\end{center}
\end{table}

For the BF measurements, most systematic uncertainties related to the tag side are canceled when the double-tag technique is employed;
therefore, systematic uncertainties arise mainly from the reconstruction of the SL decay.
The systematic uncertainty associated with the tag yield for the $D^{0}$ ($D^+$) signal is estimated to be 0.2\% (0.4\%) by varying the $M_{\rm BC}$ fit range.
The uncertainties related to the $\pi^{\pm}$ tracking efficiency, $\pi^{\pm}$ particle identification (PID) efficiency
and $\pi^0$ reconstruction efficiency are estimated to be 0.8\% (1.2\%), 0.2\% (0.3\%), and 0.6\%, respectively, 
by studying the doubly tagged $D\bar{D}$ hadronic decay samples. 
Using a sample of radiative Bhabha events, the uncertainty of the $e^{\pm}$ PID efficiency is estimated to be 0.5\% for both modes.
The uncertainty from the $e^{\pm}$ energy recovery is estimated to be 0.4\% (0.7\%) by comparing to the BFs obtained without recovery.
The uncertainty from the $K^0_S$ veto is estimated to be 1.8\% by varying the size of the veto window.
The fully reconstructed $D\bar{D}$ hadronic decays are used to show that the uncertainty due to the $E_{\gamma, \text{max}}$ requirement is negligible.
We estimate the uncertainty in the signal yield of the $U_{\text{miss}}$ fit to be 1.5\% (0.5\%) by varying the fitting range.
The uncertainty related to the modeling of the background shape is estimated to be 1.5\% (1.4\%) by changing the BFs of the dominant background channels by $\pm 1\sigma$,
and $\sigma$ is the uncertainty reported in Ref.~\cite{pdg}.
We estimate the uncertainty due to the PWA model of the signal to be 0.3\% (0.9\%)
by varying the parameters of the nominal solution by their statistical uncertainty.
These estimates are added in quadrature to obtain the total systematic uncertainty of 2.5\% (3.0\%) for $D^0$ ($D^+$) mode.

The following sources of systematic uncertainties, as summarized in Table~\ref{tab:ff-syst-sum}, have been considered in the PWA procedure.
The uncertainty related to variations to the fit are estimated by taking the difference between the alternative fit and the nominal fit.
The uncertainty from the modeling of the background shape is assigned as for the BF measurement.
The uncertainty due to the fixed background fraction $f_b$ is
estimated by changing by $\pm 1\sigma$ of its statistical error.
The parameter of $r_{BW}$ is set to 3.0 GeV$^{-1}$ in the nominal fit; 
the uncertainty related to this imperfect knowledge is estimated by varying the value within 2.0 -- 4.0 GeV$^{-1}$.
We vary $m_V$ and $m_A$ by $\pm 100\ \text{MeV}/c^2$ to estimate the uncertainties associated with the pole mass assumption.
The uncertainty from the $\rho$ or $\omega$ line shape is estimated by varying the mass and width of $\rho$ or $\omega$ by $\pm 1\sigma$ error~\cite{pdg}.
The systematic uncertainty of the $f_0(500)$ modeling is considered by replacing with a conventional RBW function with
the mass and width fixed to the BESII measurements~\cite{sigma-bes2}.
The possible bias due to the fit procedure is studied with the same method described in Ref.~\cite{bes3-luy}. 
The mean bias is taken as a corresponding systematic uncertainty.

\begin{table}[htp]
\begin{center}
\caption{Absolute systematic uncertainties on the FF ratios and the fractions of different components in \Dp decays.}
\vspace{0.25cm}
\scalebox{0.88}{
\begin{tabular}{l|c|c|c|c|c}
\hline
\hline
Source                                 & $r_V$          &  $r_2$ & $f_{f_0(500)}$ (\%) & $f_{\rho^{0}}$ (\%) & $f_{\omega}$ (\%) \\
\hline
Background shape                       & 0.003 & 0.003 & 0.06 & 0.06 & 0.009 \\
Background fraction                    & 0.008 & 0.021 & 0.32 & 0.25 & 0.060 \\
$r_{BW}$                               & 0.024 & 0.026 & 0.56 & 0.56 & 0.059 \\
$m_V$                                  & 0.035 & 0.001 & 0.02 & 0.02 & 0.004 \\
$m_A$                                  & 0.025 & 0.020 & 0.06 & 0.04 & 0.013 \\
$\rho$ line shape                      & 0.002 & 0.003 & 0.05 & 0.02 & 0.034 \\
$\omega$ line shape                    & 0.0002& 0.0002& 0.02 & 0.09 & 0.008 \\
$f_0(500)$ modeling                    & 0.012 & 0.005 & 0.83 & 0.88 & 0.038 \\
Fit procedure                          & 0.003 & 0.003 & 0.18 & 0.27 & 0.086 \\
\hline
Total                                  & 0.051   &0.039 & 1.07 & 1.11 & 0.15\\
\hline
\hline
\end{tabular}
}
\label{tab:ff-syst-sum}
\end{center}
\end{table}

In summary, the SL decays \Dosig and \Dpsig are studied using a data sample
corresponding to an integrated luminosity of 2.93~fb$^{-1}$ collected with the BESIII detector at $\sqrt{s}= 3.773$ GeV.
We measure the FF in $D\to \rho e^+\nu_e$ via a simultaneous PWA fit to both decay channels, and improve the absolute BFs for these decays. 
The FF measurements are consistent with the only measurement~\cite{CLEO-D2rhoenu} but with improved precision. 
These measurements are compatible with the theoretical calculations~\cite{LQCD, QCDSR} that have much larger uncertainty than experimental results. 
They can also aid the determination of $V_{ub}$ via a double-ratio technique~\cite{DbRatio}.
The BFs results are consistent with isospin invariance: $\frac{\Gamma(D^0\to\rho^-e^+\nu_e)}{2\Gamma(D^+\to\rho^0e^+\nu_e)} = 0.985\pm0.054\pm0.043$.
The BFs of different components contributing to the \Dpsig decay are also obtained.
The hadronic system in this decay is dominated by the $P$ wave, which is mostly a $\rho^0$ contribution along with a much smaller one from the $\omega$.
Additionally, the $S$-wave process $D^+\to f_0(500)e^+\nu_e$ is observed for the first time with a relative contribution of $(25.7\pm1.6\pm1.1)\%$.
This is compatible with the theoretical predictions reported in Refs.~\cite{Takayasu,YuJiShi}.
The process $D^+\to f_0(980)e^+\nu_e$ is not significant and an upper limit on its BF is set at the 90\% CL.

In the SU(3) symmetry limit, Ref.~\cite{wangwei} proposed a model-independent way
to distinguish the two different descriptions of the scalar mesons using a ratio
$R=\frac{\mathcal{B}(D^{+} \to f_0(980) e^+ \nu_e) + \mathcal{B}(D^{+} \to f_0(500) e^+ \nu_e) }{\mathcal{B}(D^{+} \to a_0(980)^0 e^+ \nu_e)}$,
which is predicted to be $1.0\pm0.3$ for the two-quark description and $3.0\pm0.9$ for the tetraquark description.
We obtain $R > 2.7$ at the 90\% CL by using $\mathcal{B}(f_0(500)\to\pi^+\pi^-)=67\%$,
$\mathcal{B}(a_0(980)^0\to\pi^0\eta)=85\%$~\cite{pdg} and the BESIII measurement~\cite{douzl} for $D^{+} \to a_0(980)^0 e^+ \nu_e$.
Here, we neglect the $f_0(980)$ component and assume that the dominant decays are $\pi\pi$ for $f_0(500)$, and $\pi\eta$ and $K\bar{K}$ for $a_0(980)^0$.
Our result favors the SU(3) nonet tetraquark description of the $f_0(500)$, $f_0(980)$ and $a_0(980)$.

The BESIII collaboration thanks the staff of BEPCII and the IHEP computing center for their strong support.
This work is supported in part by National Key Basic Research Program of China under Contract No. 2015CB856700;
National Natural Science Foundation of China (NSFC) under Contracts Nos. 11075174, 11121092, 11405046, 11475185, 11575091, 11625523, 11635010, 11775246;
the Chinese Academy of Sciences (CAS) Large-Scale Scientific Facility Program;
the CAS Center for Excellence in Particle Physics (CCEPP);
Joint Large-Scale Scientific Facility Funds of the NSFC and CAS under Contracts Nos. U1332201, U1532257, U1532258;
CAS under Contracts Nos. KJCX2-YW-N29, KJCX2-YW-N45, QYZDJ-SSW-SLH003;
100 Talents Program of CAS; National 1000 Talents Program of China;
INPAC and Shanghai Key Laboratory for Particle Physics and Cosmology;
German Research Foundation DFG under Contracts Nos. Collaborative Research Center CRC 1044, FOR 2359;
Istituto Nazionale di Fisica Nucleare, Italy;
Joint Large-Scale Scientific Facility Funds of the NSFC and CAS;
Koninklijke Nederlandse Akademie van Wetenschappen (KNAW) under Contract No. 530-4CDP03;
Ministry of Development of Turkey under Contract No. DPT2006K-120470;
National Natural Science Foundation of China (NSFC) under Contract No. 11505010; National Science and Technology fund;
The Swedish Research Council; U. S. Department of Energy under Contracts Nos. DE-FG02-05ER41374, DE-SC-0010118, DE-SC-0010504, DE-SC-0012069;
University of Groningen (RuG) and the Helmholtzzentrum fuer Schwerionenforschung GmbH (GSI), Darmstadt;
WCU Program of National Research Foundation of Korea under Contract No. R32-2008-000-10155-0.

\end{document}